\documentclass[sigconf]{acmart}

\usepackage{amsthm}
\usepackage{multirow}
\usepackage{enumitem}
\usepackage{todonotes}
\usepackage{listings}
\usepackage[utf8x]{inputenc}
\newcommand{\baseoness}{FakeFol\textsubscript{ss}}
\newcommand{\baseones}{FakeFol\textsubscript{s}}
\newcommand{\basetwo}{FolMarket}

\newcommand{\modelname}{\texttt{DECIFE}}

\copyrightyear{2021} 
\acmYear{2021} 
\setcopyright{acmcopyright}\acmConference[HT '21]{Proceedings of the 32nd ACM Conference on Hypertext and Social Media}{August 30-September 2, 2021}{Virtual Event, Ireland}
\acmBooktitle{Proceedings of the 32nd ACM Conference on Hypertext and Social Media (HT '21), August 30-September 2, 2021, Virtual Event, Ireland}
\acmPrice{15.00}
\acmDOI{10.1145/3465336.3475108}
\acmISBN{978-1-4503-8551-0/21/08}

\begin{document}
\fancyhead{}

\title{DECIFE: Detecting Collusive Users Involved in Blackmarket Following Services on Twitter}

\author{Hridoy Sankar Dutta}
\authornote{Both authors contributed equally.}
\email{hridoyd@iiitd.ac.in}
\affiliation{%
  \institution{IIIT-Delhi}
  \country{India}
}
\author{Kartik Aggarwal}
\authornotemark[1]
\email{kartikaggarwal98@gmail.com}
\affiliation{%
  \institution{IIIT-Delhi}
  \country{India}
}
\author{Tanmoy Chakraborty}
\email{tanmoy@iiitd.ac.in}
\affiliation{%
  \institution{IIIT-Delhi}
  \country{India}
}

\renewcommand{\shortauthors}{Trovato and Tobin, et al.}

\begin{abstract}
  The popularity of Twitter has fostered the emergence of various fraudulent user activities - one such activity is to artificially bolster the social reputation of Twitter profiles by gaining a large number of followers within a short time span. Many users want to gain followers to increase the visibility and reach of their profiles to wide audiences. This has provoked several blackmarket services to garner huge attention by providing artificial followers via the network of agreeable and compromised accounts in a collusive manner. Their activity is difficult to detect as the blackmarket services shape their behavior in such a way that users who are part of these services disguise themselves as genuine users.

In this paper, we propose \modelname, a framework to detect collusive users involved in producing `following' activities through blackmarket services with the intention to gain collusive followers in return. We first construct a heterogeneous user-tweet-topic network to leverage the follower/followee relationships and linguistic properties of a user. The heterogeneous network is then decomposed to form four different subgraphs that capture the semantic relations between the users. An attention-based subgraph aggregation network is proposed to learn and combine the node representations from each subgraph. The combined representation is finally passed on to a hypersphere learning objective to detect collusive users. Comprehensive experiments on our curated dataset are conducted to validate the effectiveness of \modelname\ by comparing it with other state-of-the-art approaches. To our knowledge, this is the first attempt to detect collusive users involved in blackmarket `following services' on Twitter.
\end{abstract}

\ccsdesc[500]{Information systems~Social networks}
\ccsdesc[500]{Security and privacy}
\keywords{Followers, collusion, blackmarket, Twitter, OSNs}

\maketitle

\section{Introduction}
Are you a Twitter user? Do you want to boost your Twitter profile (by increasing the follower count) within a limited time without getting suspended by Twitter? Several online services are ready to assist you. What you need to do is simple -- pay them, and they will provide you followers; most of these followers would be legitimate Twitter users. If you can not afford to pay money, then you can opt for another option -- just become a part of these services, start following their customers and earn credits; these credits can be used further to gain your own followers. Do not worry about being flagged by Twitter policy as these services are so smart in their following mechanism that they can easily deceive Twitter into thinking that their activity is legitimate. This is the philosophy behind many online blackmarket following services.

\begin{figure*}
    \centering
    \includegraphics[width=\textwidth]{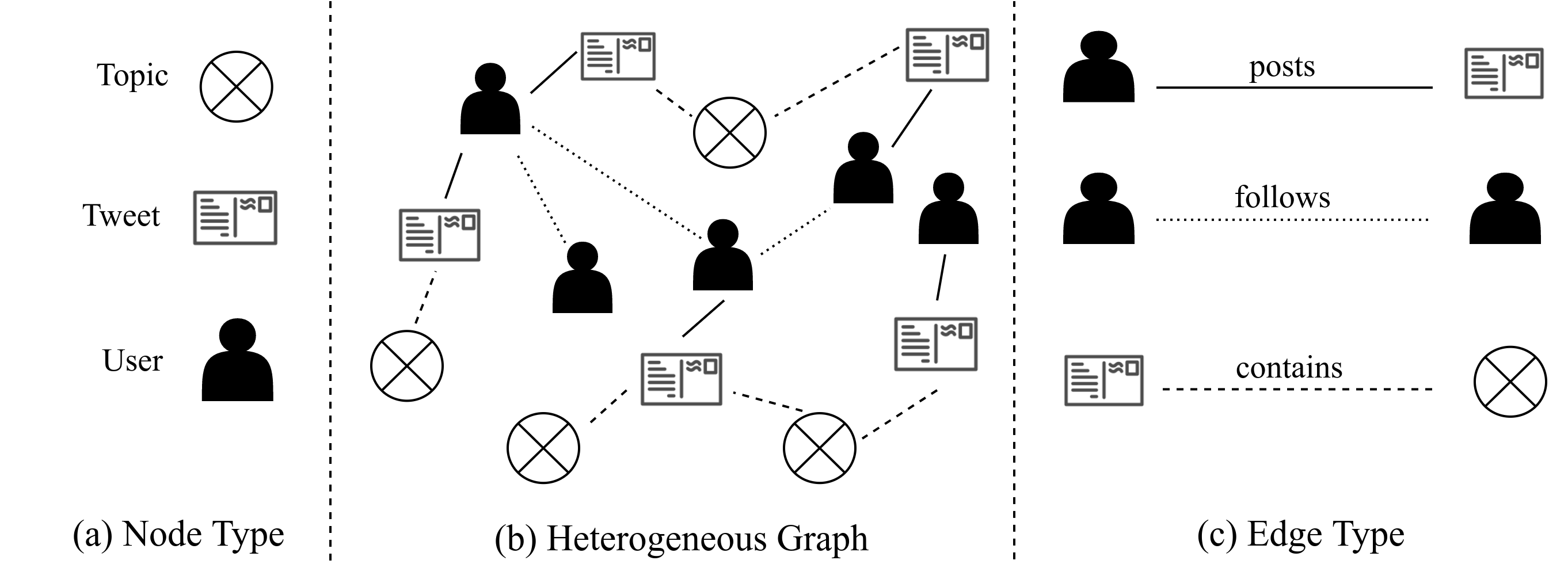}
    \caption{A heterogeneous network for modeling collusive users and their interactions. (a) Three types of nodes. (b)  User-tweet-topic heterogeneous network. (c) Three types of edges involved in the network. The detailed construction of the heterogeneous network can be found in Section \ref{sec:graph_construction}.}
    \label{fig:heterogeneous_graph}
\end{figure*}

Twitter is arguably the most popular Online Social Network (OSN) for mass communication. It is also one of the key platforms for digital campaigning, social networking and opinion dissemination. The popularity of Twitter has attracted with it new online markets that help its users to make their profiles attractive by increasing followers, retweets, likes, replies, etc. In this context, Stringhini et al.  \cite{b17} coined the term ``Twitter followers market" to characterize those syndicates catering to people willing to pay for a quick increase of their followers. Many people try to rapidly gain fame by exploiting this mechanism --  they buy followers from these online markets. 
However, the coordinators of these services manage the activities of users to maintain the integrity and avoid societal, privacy and security problems. Controlling such type of artificial following activities has thus become one of the major challenges.
There exist several blackmarket agencies which have created thriving and intelligent ecosystems of producing illicit followers. Users can gain such followers by paying money ({\em premium services}) or for free by following customers of those services ({\em freemium services}) \cite{b10}. In this paper, we focus our attention on the latter case, and call this type of users ``{\bf collusive users}'' -- \textbf{\textit{users who gain followers from blackmarket services}}. Previous literature \cite{dutta2018retweet,dutta2019blackmarket,arora2020analyzing,dutta2020hawkeseye,chetan2019corerank} on collusive entities in online media reported that collusive users are not `fake'  -- they are not bots, the handles of these Twitter profiles are normal human beings, and their following activities are not mechanically controlled by any predefined policy.
Rather, collusive users exhibit {\em a hybrid following behavior} -- on one hand, similar to non-collusive users,  they organically follow other users due to similar topical interest; on the other hand, similar to fake users, they inorganically follow other blackmarket customers only to gain credits. Designing a system to identify collusive users would be useful for Twitter managers and social network analysts to figure out how and what extent the profile of a user has turned out to be popular due to the support of such blackmarket services. 

In this work, we propose \modelname, a novel heterogeneous graph attention network for detecting collusive users who are involved in producing fake followers. We first create a heterogeneous network (c.f. Figure \ref{fig:heterogeneous_graph}) based on the relationship and linguistic properties of users. We then decompose the network into four different subgraphs to capture semantic relations between users. Finally, we exploit a hierarchical aggregation network to learn node representations that are passed on to a hypersphere learning objective to detect collusive users. To evaluate our method, we collected data of collusive users from a popular blackmarket service by designing a customized web scraper. The entire data collection was carried out after taking proper IRB (institutional review board) approval from our institute. We show the effectiveness of our proposed \modelname~ model by comparing it with three baseline methods -- \baseoness \cite{Castellini:2017}, \baseones \cite{Castellini:2017}, \basetwo~\cite{aggarwal2015} and three individual components of \modelname~considered in isolation (ablation study). 




In summary, the major contributions of the paper are four-fold:
\begin{enumerate}[leftmargin=*]
    \item We deal with a \textbf{novel problem} of detecting collusive users in Twitter where normal users get involved in artificial following activities through blackmarket services for boosting their online profiles. { \em To the best of our knowledge, no existing work has investigated the problem of `collusive user detection involved in blackmarket following services' on Twitter.}
    \item We introduce \modelname, a \textbf{novel framework} to detect collusive users on Twitter. It uses a hierarchical subgraph aggregation framework to leverage the follower/followee relationships and linguistic properties of users.
     \item We prepare \textbf{a new dataset} of collusive users who are involved in blackmarket following services on Twitter. {\em This, to our knowledge, is the first dataset of this kind.}
    \item We conduct extensive experiments on the curated dataset of collusive users to show the \textbf{superiority of our method} over state-of-the-art approaches. 
\end{enumerate}


\noindent{\underline{\textbf{Reproducibility:}}} To encourage reproducible research, we have made the codes
and the anonymized version of the dataset publicly available at \textcolor{blue}{https://github.com/LCS2-IIITD/DECIFE}.

\section{Related Work} 
We discuss the related literature by dividing the existing work into two parts -- (i) detection of fake followers in OSNs, and (ii) study of blackmarket services in OSNs.

\subsection{Detection of Fake Followers in OSNs}\label{sec:lit1}
Most of the approaches to detect fake followers identify a set of features and use machine learning techniques. Wiltshire, in her blog\footnote{\url{https://www.gshiftlabs.com/social-media-blog/the-fake-followers-epidemic}}, discussed the population of fake followers and found that there is an increase of 1-3\% fake followers for a Twitter account every couple of months. 
    \citet{b4} mentioned that there are two reasons to buy fake followers:  increase visibility and push advertisements. A tool, called ``Fake Followers Check''\footnote{\url{https://bit.ly/2KzMRdd}} was developed  to detect fake followers on Twitter based on the ratio of friends and followers, usage of repeated spam phrases, count of retweets, etc. 
    \citet{b4} proposed a machine learning approach to detect fake followers using multiple features and different machine learning classifiers. \citet{b8} used centrality-based graph features to detect fake Twitter followers. 
    \citet{tist1} found retweeters who can be used to spread the message effectively among different group of people.  
    \citet{b11} proposed fBox, an algorithm to detect suspicious friend and follower links on a large who-follows-whom Twitter dataset. \citet{b12} developed a classifier for fake follower detection using profile, timeline and relationship based features. \citet{b13} proposed CatchSync to detect suspicious nodes (followers and botnets) exhibiting synchronized behavior on Twitter social network. \citet{b26} studied the dynamics of unfollow behavior in Twitter based on online relationships of Korean-speaking Twitter users.  
    \citet{sac1} identified users with increased follower count using unsupervised local neighborhood detection method. 
    \citet{nisc1} proposed supervised spammer detection method with social interaction to detect spammers on twitter based on content and social interaction.  
    \citet{icdm14} detected campaign promoters by mapping the problem to relational classification and solved it using typed Markov Random Fields. 
    \citet{commun1} discussed how social bots which interact with humans and get unnoticed have risen in the present scenario.
\citet{b5} distinguished fake followers from the legitimate users using several discriminative features. \citet{b6} used a network-based strategy to identify fake followers.
\citet{b14} focused on the detection of zombie followers in Sina Weibo.

\subsection{Study of Blackmarket Services in OSNs}\label{sec:lit2}
 Blackmarket services have gained substantial attention recently because of the techniques they use to provide services to the customers. 
 \citet{b17} was the first to analyze the Twitter follower markets based on the market size and market price. 
 A detailed analysis of blackmarket services is presented in \cite{b15,b16} with the impact on multiple OSNs. Most of the prior studies showed how fake followers in social media help in promoting different agenda \cite{b17,b18,aggarwal2015}. \citet{farooqi2017measuring} showed how collusion networks collect `OAuth' access tokens from colluding members and abuse them to provide fake likes or comments to their members. \citet{zhu2016new} proposed an automated approach to detect collusive behavior in question-answering systems. \citet{weerasinghe2020pod} studied models for detecting Instagram posts that gained interaction through collusive networks. \citet{aggarwal2015} also discovered an oligopoly structure of merchants involved in blackmarket services. \citet{b10} studied multiple types of blackmarket agencies and analyzed a honeypot fraudster ecosystem to provide insights about multifaceted behaviour of different fraudsters. \citet{b19} analyzed structure of social networks present on six different underground forums to understand the social dynamics of e-crime markets. \citet{b20} studied various blackmarkets and developed a classifier to detect fraudulent accounts sold by these marketplaces. \citet{b22} studied the behavioral characteristics of Twitter follower market merchants based on user and content based features. \citet{liu2016pay} detected `volowers' (followers who provide voluntary following services) who make profit in the follower markets. Recently, there are some preliminary works of collusive user detection on Twitter and YouTube. \citet{dutta2018retweet,dutta2019blackmarket,arora2020analyzing} proposed techniques to detect collusive users involved in blackmarket-based retweeting service on Twitter. \citet{dutta2020detecting} proposed \texttt{CollATe}, an end-to-end framework to detect collusive entities on YouTube fostered by various blackmarket services. We encourage the readers to go through \cite{dutta2020blackmarket} for a comprehensive survey on collusive activities in different online media platforms.  \\

\noindent\textbf{\large{\underline{Differences with Previous Studies:}}} The fundamental differences between the studies discussed above on fake follower detection and the collusive user detection are two-fold: (i) unlike fake followers who are mostly bots or whose activities are  mechanically controlled by predefined policies, collusive users are normal human beings, and they themselves control their accounts. Therefore, unlike fake users who mostly show ``synchronous behavior'' \cite{b13}, {\em collusive users are asynchronous in nature} \cite{dutta2018retweet}. (ii) Unlike fake followers, collusive users exhibit a hybrid following behavior -- in one hand, being a normal user, they follow other users organically due to similar topical interest; on the other hand, being a collusive user, they randomly follow other blackmarket customers inorganically to gain credits \cite{dutta2019blackmarket}. 
\textit{To the best of our knowledge, ours is the first work to detect \textbf{collusive users who gain artificial followers} from blackmarkets}.   

\section{Background and Dataset} \label{sec:dataset}
Approaching blackmarket services is one of the quickest ways to boost the impact of users on social media. These blackmarket services offer promotional services on multiple online media such as OSNs (e.g., likes on Facebook; followers, retweets, likes on Twitter;  followers on Instagram), subscription-sharing platforms (e.g., views, subscribers, likes on Youtube), music-sharing platforms (e.g., plays, followers, likes, reposts, comments on SoundCloud,  fans on ReverbNation), business and employment-oriented platforms (e.g., followers, connections, endorsements on LinkedIn), etc. 
We identified the blackmarket services providing collusive follower appraisals for Twitter by querying on search engines with keywords such as ``buy free followers'', ``get me followers'', ``get followers quickly''. 
To collect collusive users, we selected YouLikeHits\footnote{https://www.youlikehits.com/twitter2.php}, a popular credit-based freemium blackmarket service\footnote{Though there exist several blackmarket services to get collusive appraisals, we choosed only YouLikeHits for our study due to its popularity and extensive amount of literature \cite{dutta2018retweet,dutta2019blackmarket,dutta2020detecting} investigating this service.}. Customers on navigating to these services can see a dashboard (earning area) of other customers who are also using that service. YouLikeHits provide the customers with an initial credit of $50$ which can be utilized to use various facilities offered by the service. After taking proper IRB approval from our institute, we developed a web scraper that used Selenium\footnote{\url{https://www.seleniumhq.org/}} to start a headless web browser navigating to the URL of the blackmarket service. We used popular Python packages such as Requests, BeautifulSoup etc. to parse the Twitter user IDs who submitted their profile for collusive follower appraisals. We further used the Tweepy library\footnote{\url{https://github.com/tweepy/tweepy}} to collect the metadata and timeline of the Twitter users. Note that we also collect a set of non-collusive users who surely have not participated in any kind of blackmarket-driven activities to gain artificial appraisals. This set of users is only collected for the test phase of our experiment and is not the part of original dataset.

\begin{figure*}
    \centering
    \includegraphics[width=\textwidth ]{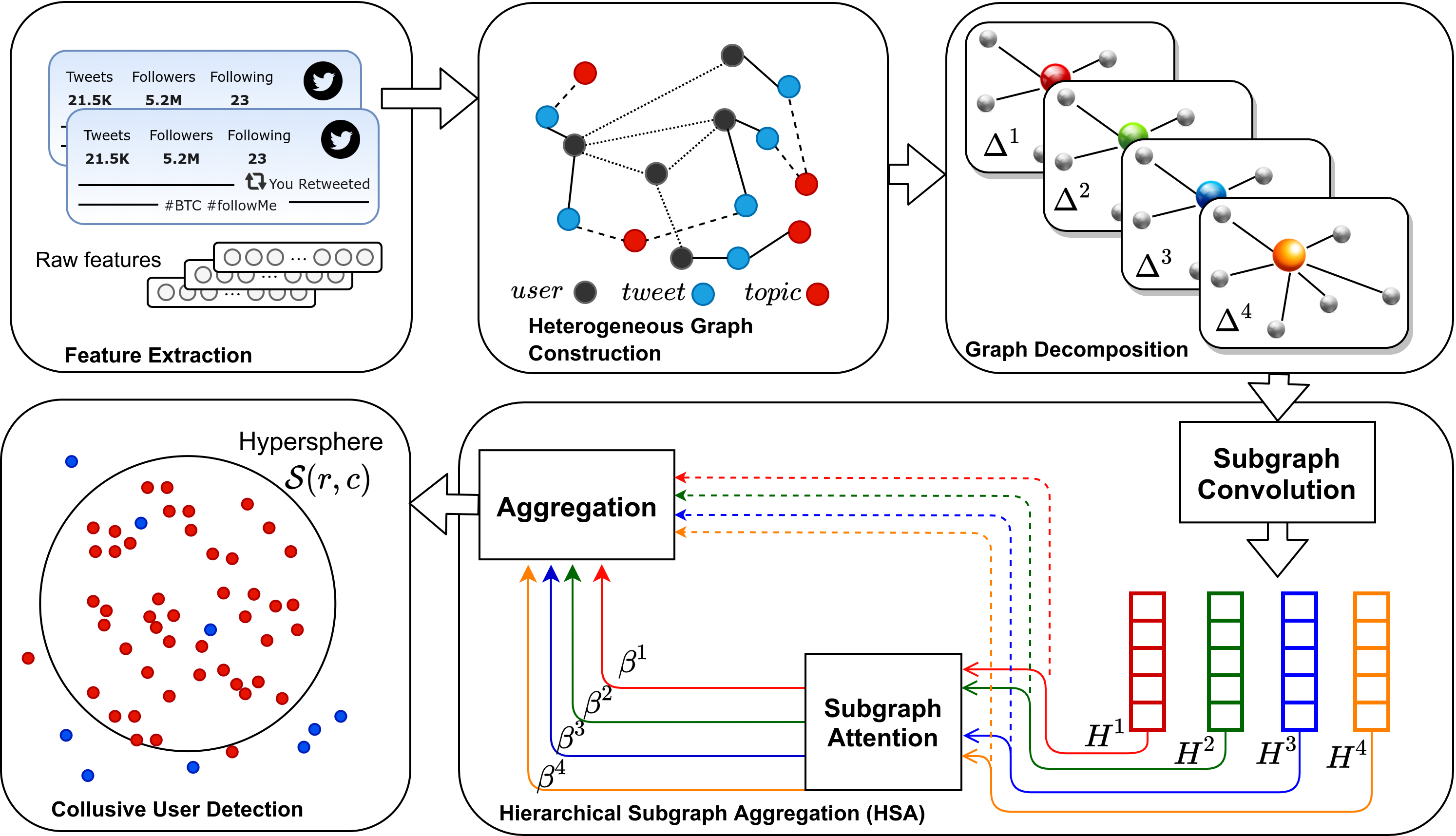}
    \caption{Architecture of our proposed \modelname\ model for collusive user detection on Twitter.}
    \label{fig:my_label}
\end{figure*}

\section{\textbf{\modelname:} Our Proposed Model}

 The objective of collusive user detection task is to learn how to automatically identify Twitter users who submitted their accounts to blackmarket services to gain collusive followers. Formally, given a set of collusive users $u_i = \{u_1, u_2, \dots \}$, connected via relationships $\Delta_m = \{\Delta_1, \Delta_2, \dots \}$ in a heterogeneous network $\mathcal{G}$, \modelname\ learns a hypersphere boundary $\mathcal{S}(r,c)$ (with radius $r$ and center $c$) around the collusive users to identify whether a new user is a collusive user or not. In this section, we present the architecture of \modelname. 
It consists of four major components: \textit{network construction}, \textit{feature extraction}, \textit{hierarchical subgraph aggregation} and \textit{collusive user detection}. In this section, we introduce each of these components in detail. Figure \ref{fig:my_label} shows the schematic architecture of \modelname.

\subsection{Network Construction} \label{sec:graph_construction}
Here, we show the construction of the heterogeneous network and subsequent subgraphs for modeling the interactions of Twitter users.

\textbf{Heterogeneous Network: } To model the Twitter network, we construct a heterogeneous network \cite{sun2013mining} as a directed network $\mathcal{G}=\mathcal{(V,E)}$, where each node $v\in \mathcal{V}$ and each edge $e\in \mathcal{E}$ are related with their node-type mapping functions
$\rho(v):\mathcal{V}\rightarrow \mathcal{X}$ and edge-type mapping functions $\varphi(e): \mathcal{E}\rightarrow \mathcal{Y}$, respectively. 
$\mathcal{X}=$ \{user($U$), tweet($T$), topic($O$)\} denote the node-types and $\mathcal{Y} =$ \{follows($r$), posts($p$), contains($c$)\} denote the edge-types such that $|\mathcal{X}|+|\mathcal{Y}| > 2$ (a heterogeneous network contains at least 2 node-types or edge-types). An illustration of this network is shown in Figure \ref{fig:heterogeneous_graph}. Every Twitter user $u_i\in U$ is represented with a feature vector $x_i$ in   $\mathcal{G}$ and is connected with its corresponding tweets $t_{ij}\in T$ and to users she follows or is followed by. Every tweet $t_{ij}$ of a user $u_i$ is connected to a topic node $o_k\in O$ via one-to-one mapping, where $k$ is the set of topics.


\textbf{Network Decomposition:} The heterogeneous network is decomposed into subgraphs based on multiple relationships. A Relationship $\Delta$ between a labelled\footnote{We refer to a labelled user as a user whose ground-truth label is known. In our case, it refers to a collusive user which we collected from the blackmarket service.} user-node $u_i$ and another labelled user-node $u_j$ is a sequence of connected relations $\mathcal{M}_1\rightleftharpoons\mathcal{M}_2\dots \rightleftharpoons\mathcal{M}_n$ in   $\mathcal{G}$. Each relation $\mathcal{M}$ from source node $v_1\in \mathcal{V}$ to target node $v_2\in \mathcal{V}$ with edge $e\in \mathcal{E}$ is denoted as $\langle \mathcal{X}(v_1), \mathcal{Y}(e), \mathcal{X}(v_2)\rangle$. Note that $(v_1,v_2)$ can belong to any node-type connected via edge $e$.

To capture different aspects of a user's behavior, we use four types of relationships, $\Delta_m, m=\{1,2,3,4\}$ which are detailed below:
\begin{enumerate}
    \item $\Delta_1$: To describe a connection between users in the `following' blackmarket services, we formulate a relationship having a \textbf{common follower}.
    \begin{center}
        user$_1\xrightarrow[]{follows}$user$_x\xleftarrow[]{follows}$user$_2$
    \end{center}
    \item $\Delta_2$: Users in a blackmarket service tend to participate in a \textit{barter system} where they follow other users to gain credits that can later be  used to obtain followers of their accounts. As an example, a \textbf{transition user} (user$_x$) signed up on the service, \textbf{follows users} (user$_1$) to gain credits which can be availed \textbf{to gain followers} (user$_2$) in return. To exploit this characteristic, we formulate a relationship between users in a 1-hop manner.
    \begin{center}
        user$_1\xleftarrow[]{follows}$user$_x\xleftarrow[]{follows}$user$_2$
    \end{center}
    \item $\Delta_3$: To describe a \textbf{direct connection} between two users, we formulate a relationship in a 0-hop manner.
    \begin{center}
        user$_1\xrightarrow[]{follows}$user$_2$
    \end{center}
    \item $\Delta_4$: Twitter users involved in the promotion of campaigns/ websites/ events try to gain popularity from a particular target audience by gaining more retweets/followers to their tweets/accounts. To capture the connection between users tweeting/retweeting on the \textbf{same topic}, we formulate a relationship based on their common topic of interest. We detail this strategy of connecting users based on topical interest in Section \ref{sec:implementation_details}.
    \begin{center}
        user$_1\xrightarrow[]{posts}$tweet$_1\xrightarrow[]{contains}$topic$_1\xleftarrow[]{contains}$tweet$_2\xleftarrow[]{posts}$user$_2$
    \end{center}
\end{enumerate}

\subsection{Features Extraction}
In a Twitter user network, the behaviours and attributes of an account are directly linked with the user's characteristics. Taking this into account, we consider 18 features which are directly extracted or calculated from the user metadata. We describe each feature, its corresponding description and type in Table \ref{tab:meta_feature}. We then concatenate these features to form an 18-dimension user-metadata feature vector that is used to train our \modelname\ model. The user-metadata feature vectors, along with subgraph information, are fused at 2-levels using Hierarchical-Subgraph Aggregation (HSA) network to obtain final user representations.

\begin{table*}[!htbp]
    \centering
    \caption{User metadata features used by \modelname.}
    \label{tab:meta_feature}
    \scalebox{0.8}{
    \begin{tabular}{l|p{0.7\textwidth}|l}
    \hline
    {\bf Feature} & {\bf Description} & {\bf Type}\\\hline
  Follower count & Number of followers of a user & Real \\
    Friend count & Number of friends of a user & Real\\
    Status count & Number of tweets posted and retweeted by a user & Real \\
    Favorite count &  Number of status liked by a user & Real \\
    Description presence & If the account description is present & Binary\\
    Description length & Number of characters present in the user description & Real\\
    URL  & If a URL is present in the user description & Binary \\
    Location  & If the location is present in the user info & Binary\\
    Profile image & If a profile image is present in the user account & Binary\\
  Background image & If a background image is present in the user account & Binary\\
    Account age & Number of days elapsed since the account was created & Real \\
    Account entropy & Skewness in user's activity, $\sum{c_i\log(c_i)}$, where $c_i$ is the year-wise count of tweets by the user & Real\\
    Emojis & Avg. number of emojis per tweet posted by a user & Real\\
    Retweets & Ratio of retweet count to tweet+retweet count & Real\\
    URLs & Avg. number of URLs per tweet by a user & Real\\
    User mentions & Avg. number of user mentions per tweet by a user & Real\\
    Words & Avg. number of words in a tweet by a user & Real\\
    Hashtags & Avg. number of hashtags in a tweet by a user & Real\\\hline
    \end{tabular}}
\end{table*}

\subsection{Hierarchical Subgraph Aggregation (HSA)}
We employ a two-level hierarchical subgraph aggregation mechanism to detect whether a user is collusive or not. In the first step, a weighted subgraph convolution network is used to obtain hidden representations of the labelled users based on the connections of a user node $u_i$ with its associated users $u_j, j\in \mathcal{N}_i^m$ ($\mathcal{N}^m_i$ being the set of neighbors of users $u_i$ connected via relationship $\Delta_m$). Next, the subgraph attention network learns the importance of different edge-types to aggregate information among all nodes in the network.

\subsubsection{Subgraph Convolution Network: }
Considering that a pair of users in a subgraph can be connected via multiple edges (e.g., two users posting multiple different tweets on the same topic), we propose a weighted subgraph convolution network with the following layer propagation rule:
\begin{equation}
    h_i^m = \sigma(b^m + \sum_{j\in\mathcal{N}_i}\frac{e^m_{ij}}{c^m_{ij}}x_j w^m)
\end{equation}
where $e^m_{ij}$ is the edge-weight coefficient for the edge $e_{ij}$ connected via relationship $\Delta_m$ (c.f. Table \ref{tab:subgraph}), and $\mathcal{N}_i$ is the set of neighbors of  node $i$. Further, $c^m_{ij} = \sqrt{|\mathcal{N}_i||\mathcal{N}_j|}$ is the normalization term, and $|\mathcal{N}_i|$ is the degree of node $i$.
Here, $x_j$ is the input feature vector of node $j$, and $w^m$ is the learnable weight parameter for the subgraph $m$. The weight coefficient in subgraph convolution network makes it superior for learning the connectivity information between the user-nodes than a conventional graph convolution network.

In generalised-matrix form, the equation can be written as:
\begin{equation}
    H^{m} = \sigma((D^m)^{-1}A^m K^m X W^m)
\end{equation}
where $A^m$ is the adjacency matrix, $K^m$ is the edge-weight matrix, $D^m_{ii}=\sum_{j\in\mathcal{N}_i} A^m_{ij} $ is the diagonal matrix, and $W^m$ is the learnable layer-wise weight matrix of a subgraph having relationship $\Delta_m$.

Therefore, we obtain 4 groups of subgraph-specific node-embeddings $H^m$ for each given relationship $\Delta_m, m=\{1,2,3,4\}$. These are then used to compute importance among subgraphs using Subgraph Attention Network as described below.

\subsubsection{Subgraph Attention Network:} Subgraph convolution network learns a node representation by leveraging the importance between users only from a single type of relationship. However, attending to multiple relationships at once can reveal the importance of different edge-types to a node pair and aggregate rich semantic information from them. The subgraph-specific node-embeddings $H^m$ obtained from subgraph convolution network are projected using a multilayer feed-forward neural network to undergo a non-linear transformation.
\begin{equation}
    \tilde{H}^m=W_1\cdot H^m_i+b_1
\end{equation}
\begin{equation}
    w^m = \frac{1}{n}\sum_i^n W_2(\text{tanh}(\tilde{H}^m))+b_2
\end{equation}
where $W_1,W_2$ are the weight matrices and $b_1,b_2$ are the bias vectors.

The weight coefficient $\beta^m$ of each subgraph is calculated by normalizing the importance $w^m$ using a softmax function.
\begin{equation}
    \beta^m = \frac{\text{exp}(w^m)}{\sum_{k=1}^n \text{exp}(w_k)}
\end{equation}

The information from both the levels can be aggregated to obtain the final embeddings as follows:
\begin{equation}
    Z= \sum_{m=1}^3 \beta^m\cdot H^m
\end{equation}

To improve the stability of the training process, we extend this to multiple heads $d=\{d_1,d_2,\dots, d_l\}$ by passing $Z^{l-1}$ as an input through the Hierarchical Subgraph Aggregation Network,

\begin{equation}
    Z^{l} = HSA(Z^{l-1})   
\end{equation}

The final embedding $Z^l$ of each user-node is used to learn a hypersphere boundary around the collusive distribution.

\subsection{Collusive User Detection}
Given a training set $\{Z_n = Z_1,Z_2,..,Z_n\} \in \mathbf{R}^{n\times d}$ with embeddings of dimension $d$ belonging to $n$ collusive users $u_i\in U$ derived by hierarchical aggregation, our \modelname~model is trained to learn a hypersphere boundary that encloses the network representations around majority of the collusive data. The objective is to obtain a minimum volume hypersphere $S(r,c)$ with radius $r>0$ and center $c\in\mathcal{F}_k$ using a mapping function $\phi_k: Z\rightarrow \mathcal{F}_k$ to map the feature space $Z$ into Hilbert space $\mathcal{F}_k$. The loss $\mathcal{L}$ is similar to the DeepSVDD \cite{ruff2018deep}  and is defined as follows:
\begin{equation}
    \mathcal{L} = \enskip r^2 + \frac{1}{\mu n}\sum_{i=1}^n{\mathcal{\xi}_i} \\\hspace{4mm}
    s.t. \quad {||\phi_k(x_i)-c||}^2_{\mathcal{F}_k} \leq r^2 + \mathcal{\xi}_i,\quad\mathcal{\xi}_i \geq 0\enskip \forall\enskip i
\end{equation}
The slack variables $\mu\in (0,1]$ regulates the trade-off between penalties $\mathcal{\xi}_i$ and the volume of sphere $S(r,c)$, and $\mathcal{\xi}_i$ allows a smooth boundary of the sphere. The initial radius is set as 0, and the center is obtained through the first forward pass to our network model.
\begin{equation}
    c_0 = \frac{1}{n}\sum \modelname (\mathcal{G})
\end{equation}

During validation and testing, a point is marked as collusive user if it falls inside the hypersphere  $S$ such that  \small{$\quad {||\phi_k(x_i)-c||}^2_{\mathcal{F}_k} > r^2 $}.

\section{Experimental Setup}
We conduct an extensive evaluation of \modelname~on our collected dataset (mentioned in Section \ref{sec:dataset}). In this section, we start by briefly describing the baseline methods, followed by detailed evaluation. 

\subsection{Baseline Methods}\label{sec:baseline}
We consider three state-of-the-art methods  -- first two focus on fake follower detection and the third one focuses on Twitter blackmarket follower services. However, since these methods are not focused towards a network-based detection approach, we conduct an ablation study by considering one relationship at a time in our model. We also add these methods for comparing with \modelname, which is a combination of all the relationships.

\textbf{\baseoness}: Castellini et al. \cite{Castellini:2017} detected fake Twitter followers in a {\em semi-supervised} fashion by running denoising autoencoder. Their model is trained on the data of only one class (5\% of collusive users). 

\textbf{\baseones}: Castellini et al. \cite{Castellini:2017} further introduced a supervised approach to compare it with semi-supervised approach mentioned in \baseoness. They used a bunch of features related to user profile and user activity, and showed Random Forest to be the best classifier.

\textbf{\basetwo}: Aggarwal et al. \cite{aggarwal2015} characterized underground blackmarket services which provide Twitter followers. They explored three units of blackmarket services  -- merchants, customers and phony followers. They also built a supervised classifier (SVM with RBF kernel) to classify non-collusive users and phony followers.

\textbf{Variants of \modelname}:   We also derive multiple variants of \modelname\ to comprehensively compare and analyze the performances of each component. These variants are \textbf{\modelname$_{\Delta_1}$}, \textbf{\modelname$_{\Delta_2}$}, \textbf{\modelname$_{\Delta_3}$} and \textbf{\modelname$_{\Delta_4}$}. Note that $\Delta_i$ refers to relationships mentioned in Section \ref{sec:graph_construction}.

\subsection{Implementation Details}\label{sec:implementation_details}
\subsubsection{Sentence Embeddings:} Since Twitter is a global social-media platform and the intention of the collusive users is to earn followers regardless of the community, language identification of all tweets using Langid \cite{lui2012langid} revealed a mix of languages in the user timelines. Therefore, to capture the semantics of the tweets in different languages, we use Language Agnostic Sentence Embedding representations (LASER) \cite{artetxe2019massively} to build cross-lingual features of the users' tweets. For each user with $\delta$ historic tweets, we obtain a $\delta\times1024$ dimensional embeddings. These multilingual embeddings are finally passed through a cosine-similarity based K-means clustering algorithm\footnote{https://github.com/facebookresearch/faiss} for topic identification of each tweet.
\subsubsection{Parameter Settings:} We implement our model using Python 3.8.3, PyTorch 1.4.0 and Deep Graph Library \cite{wang2019deep}. The hyperparameters are tuned based on the validation set and results are reported on the test set. The penalty parameter $\mu$ is set to 0.2. The network parameters are randomly initialized with a fixed seed for all experiments and optimized using Adam optimizer. We set the learning rate to 0.6 and weight decay regularization with $\lambda= 0.0005$. For the subgraph convolution network, hidden size $H^m=32$ is used. Further, the hidden dimension $\tilde{H}^m$ for subgraph attention network is set to 128, and 2 heads are used to train the HSA model.

\subsubsection{Data Statistics:} We start by collecting a real-world Twitter user dataset for the detection of collusive users. A total of 11,076 users involved in the ``following'' activity from blackmarket service are identified and labelled as collusive users. Among these, 10,863 users are randomly used for training  \modelname, and the remaining 213 users are used for testing. The entire data collection process from the blackmarket service is mentioned in Section \ref{sec:dataset}. Along with this, we take into account 213 users which are labelled as non-collusive. \textbf{We would again like to emphasize on the fact that we chose only those Twitter users as non-collusive who are guaranteed not to be a customer of any blackmarket service. The set of non-collusive users is only used at test time to measure the performance of \modelname.}

\begin{table*}[!htbp]

\begin{minipage}{.3\linewidth}
\centering
\caption{Dataset Statistics.}\label{table:statistics}
\begin{tabular}{|c|c|}\hline
Statistic & Value\\
\hline

\# of train users & 10,863\\
\# of test users & 426\\\hline
\# of tweets & 6,627,277\\
\# of topics & 1,000\\
\hline
\end{tabular}
    \end{minipage}
    \begin{minipage}{.48\linewidth}
\centering
\caption{Subgraph Statistics.}\label{tab:subgraph}
\begin{tabular}{|c|c|c|c|}\hline
Relationship $\Delta$ & \# edges $(i,j)$ & Edge-weight $(e_{ij})$ \\
\hline
$\Delta_1$ & 6,408,903 & \# common followers \\
$\Delta_2$ & 7,206,330 & \# common transition users\\
$\Delta_3$ & 64,669 & 1\\
$\Delta_4$ & 88,069,870 & $\sum_{k\in K}\textit{min}((o_k)^i,(o_k)^j)$ \\
\hline
\end{tabular}
\end{minipage}%
\end{table*}

Further, the followers, followees and tweets of all users are extracted using the Twitter API, which are used for building the heterogeneous  network. The heterogeneous network contains around 7.9 million user nodes, 6.6 million tweet nodes and 1000 topic nodes. This network is decomposed into four subgraphs, each having $11,289$ users as nodes connected via different relationships $\Delta$.  The results are reported after 10-fold cross-validation. The collusive users are randomly partitioned into 10 groups. Each group with the same set of non-collusive users are used for each fold. Detailed statistics of the dataset and the subgraphs are shown in Tables \ref{table:statistics} and \ref{tab:subgraph}, respectively.



\section{Experimental Results and Analysis}

\subsection{Performance Comparison} In order to evaluate the performance of \modelname\ against the baseline methods, we use three popular evaluation metrics: AUC-ROC, AUC-PR and F1-score. Table \ref{table:performance} summarizes the comparative analysis of the competing methods. In general, \modelname\ outperforms all the baselines. Among the baselines, \baseones\ turns out to be the best in terms of AUC-ROC and F1-score. Among different variants of \modelname, we observe that \modelname$_{\Delta_2}$ outperforms all the other variants and our proposed model in terms of F1-score. The reason behind this is the correct identification of a higher number of non-collusive users (true positives) in comparison to the collusive users (true negatives), which is not the primary objective of our predictive modeling task\footnote{93.4\% of the non-collusive users and 60.5\% collusive users are predicted correctly by \modelname$_{\Delta_2}$.}. Note that for each of the evaluation metrics, we average the final values over 10 runs. As can be seen in Table \ref{table:performance}, our proposed \modelname\ model  consistently achieves the best performance over  others by a significant margin. We observe that the performances of the baseline methods are poor, indicating that they fail to generalize for collusive user detection as they are primarily targeted towards fake user detection. On the contrary, \modelname\ leverages the benefit of multiple relationships that prove the effectiveness of heterogeneous network to integrate multi-type information.

\begin{table}[!htbp]
\captionof{table}{Performance comparison of the competing models.}
\centering
\begin{tabular}{|l||c|c|c|}
\hline
{\bf Method} & {\bf AUC-ROC} & {\bf AUC-PR} & {\bf F1-score}  \\\hline
\baseoness & 0.417  & 0.455 & 0.537   \\
\baseones & 0.617  & 0.568 & 0.683  \\
\basetwo & 0.500 & 0.476 & 0.645 \\\hline
\modelname$_{\Delta_1}$ & 0.427 & 0.423 & 0.606 \\
\modelname$_{\Delta_2}$ & 0.689 & 0.563 & \textbf{0.802} \\
\modelname$_{\Delta_3}$& 0.818 & 0.833 & 0.745 \\
\modelname$_{\Delta_4}$ & 0.771 & 0.745 & 0.688 \\ \hline
\modelname & \textbf{0.895}
 & \textbf{0.854} & \textbf{0.786} \\\hline
\end{tabular}
\label{table:performance}
\end{table}
\hfill


\begin{figure*}
    \centering
    \includegraphics[width=0.75\textwidth]{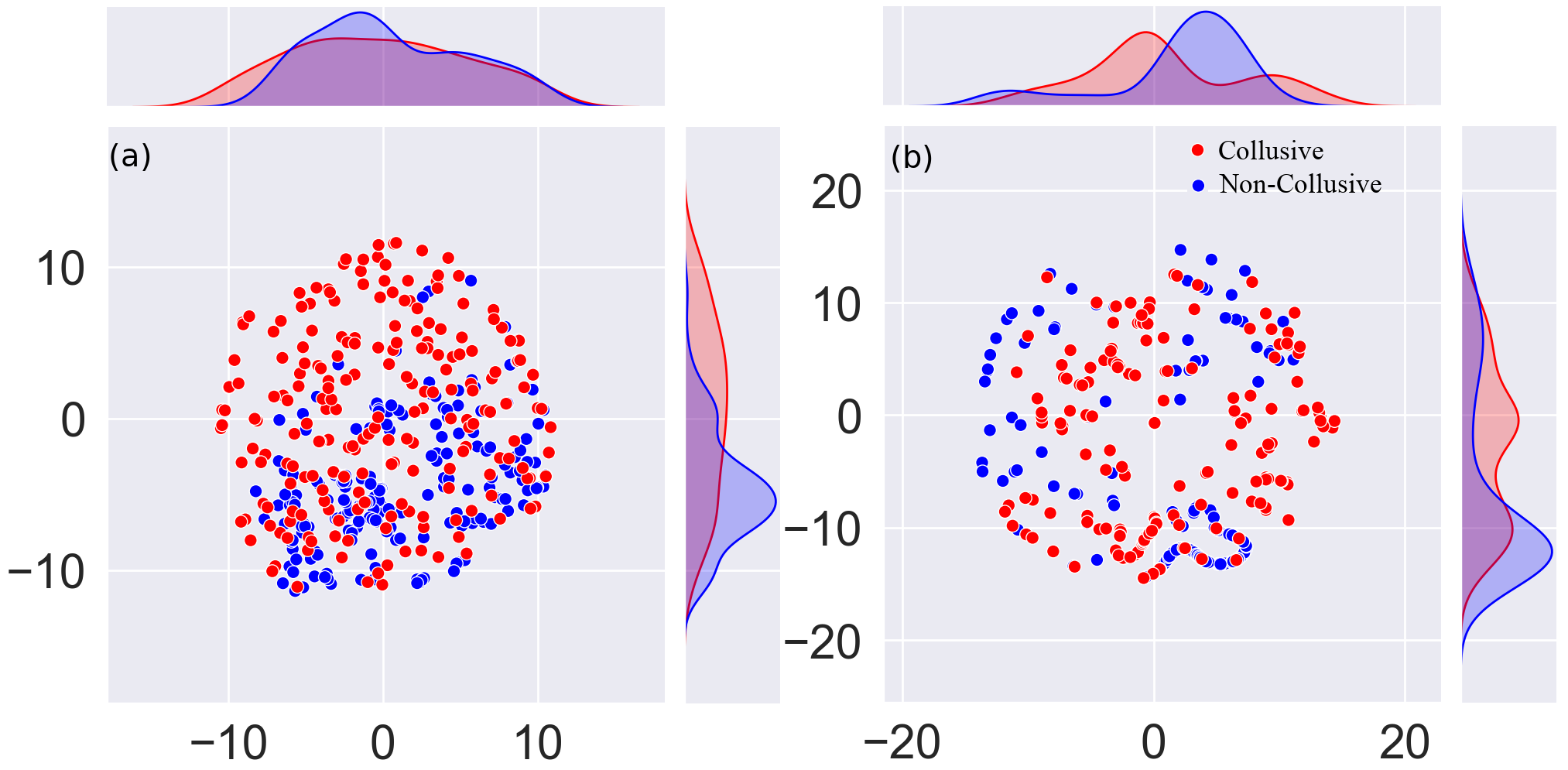}
    \caption{TSNE visualized features of collusive (red) and non-collusive (blue) users from the test dataset. The density estimate plots along the axis depict the probability distribution of the users in both sets as (a) raw features, and  (b) fully-trained model.}
    \label{fig:tsne_plot}
\end{figure*}

 The TSNE visualizations (c.f. Figure \ref{fig:tsne_plot}) of the test set embeddings for collusive and non-collusive users indicate that raw features cannot be used for detecting collusive users, whereas in the trained model, the learned hypersphere clearly encloses and distinguishes the collusive users. Moreover, the Kernel Density Estimate (KDE) plots in Figure \ref{fig:tsne_plot} depict the probability distribution of users in both sets. The probability density of collusive users in the trained model is high at points where density of non-collusive users is low, which is not the case in raw features. This suggests that the collusive users can be easily partitioned in the embedding space using the final trained embeddings by our converged \modelname\ model.

\subsection{Sensitivity of Parameters} 
\begin{figure*}
    \centering
    \begin{minipage}{0.33\textwidth}
        \centering
        \includegraphics[width=\textwidth]{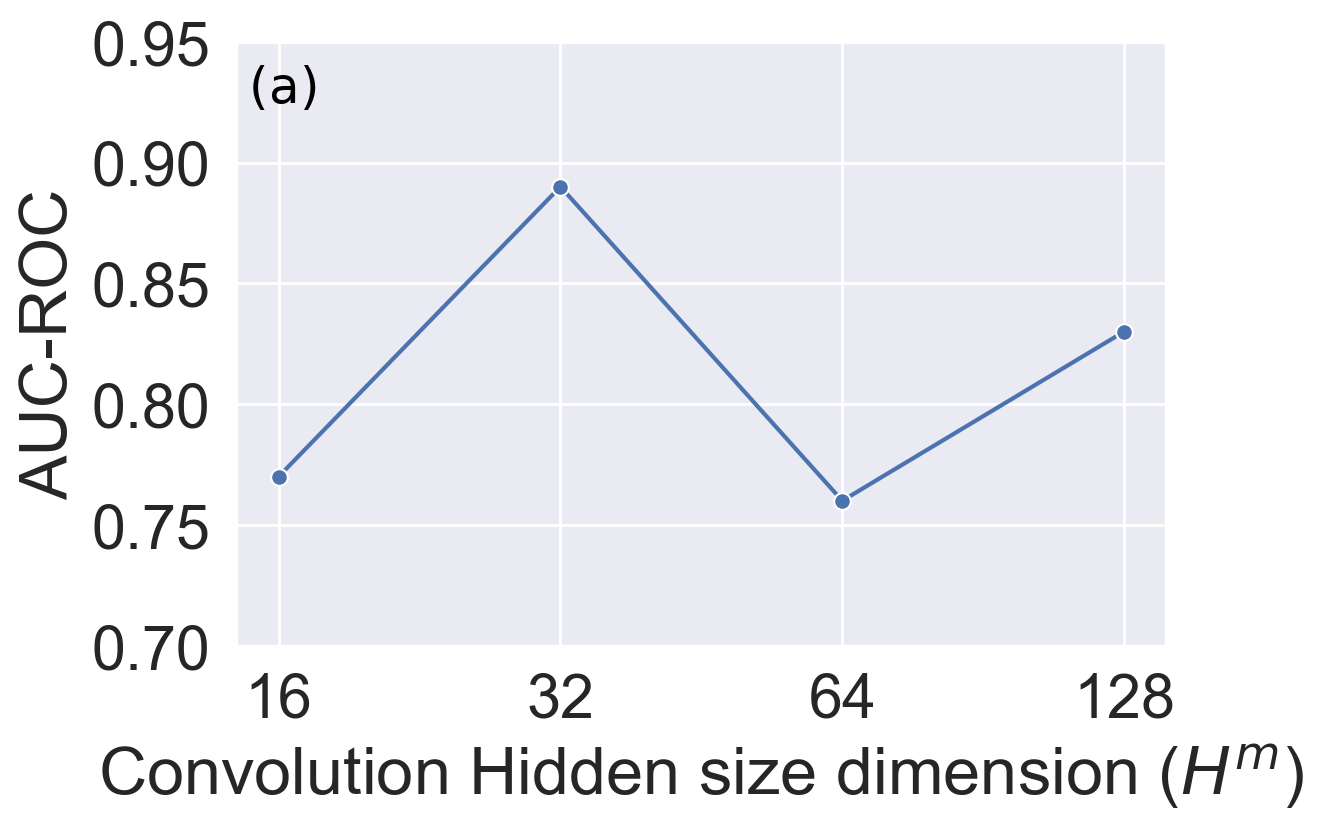} 
    \end{minipage}\hfill
    \begin{minipage}{0.33\textwidth}
        \centering
        \includegraphics[width=\textwidth]{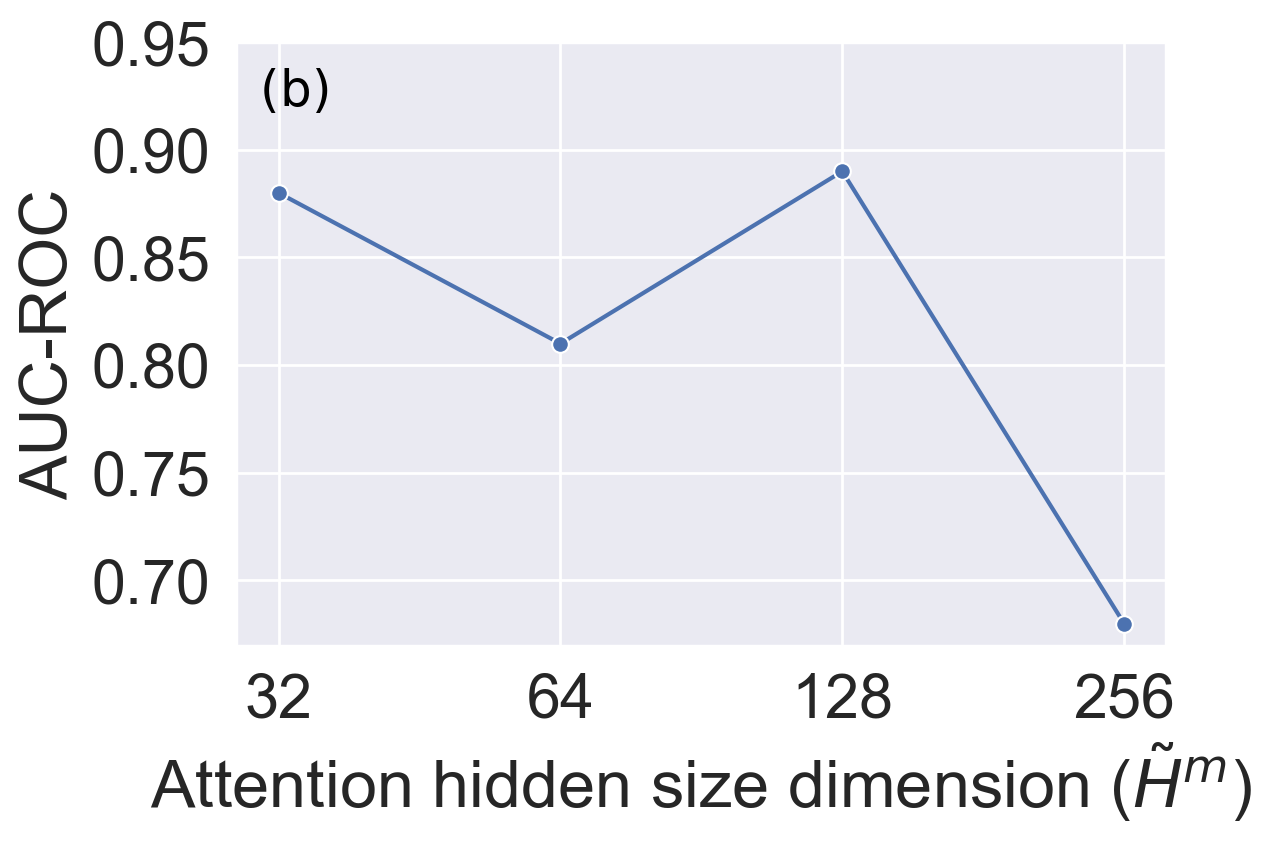} 
    \end{minipage}
    \begin{minipage}{0.33\textwidth}
        \centering
        \includegraphics[width=\textwidth]{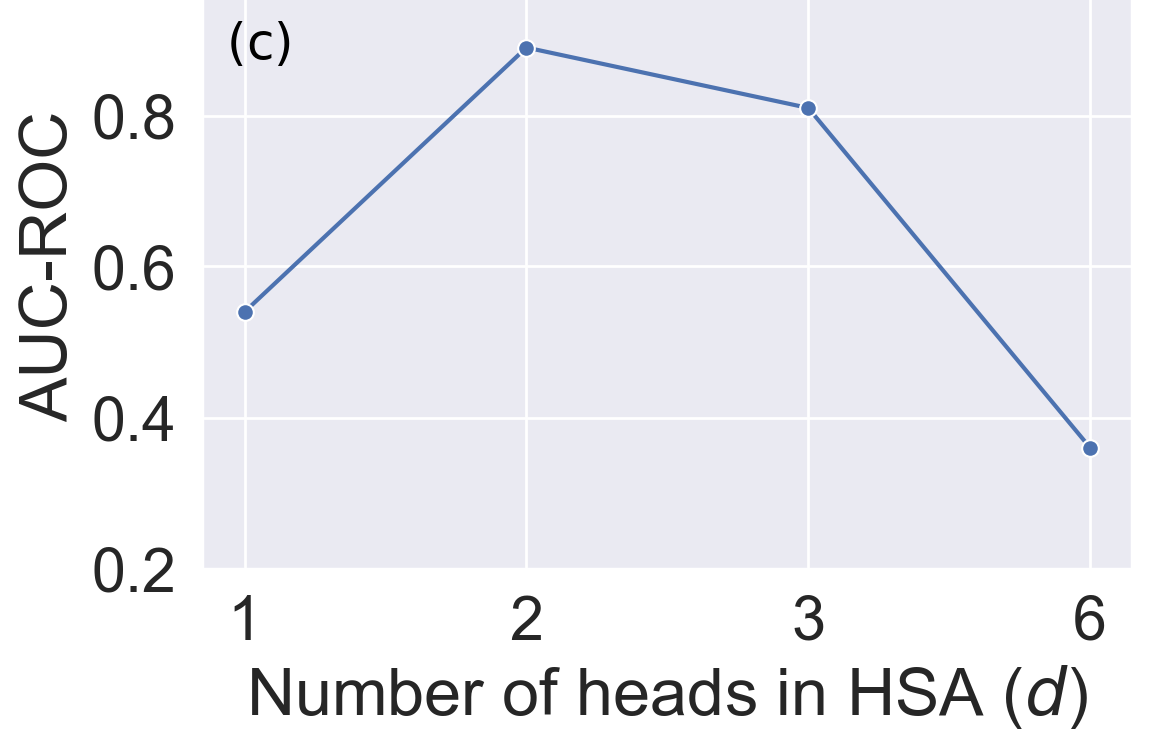} 
    \end{minipage}
    
        \caption{Parameter sensitivity of \modelname\ w.r.t (a) convolution hidden size dimension, (b) attention hidden size dimension, and (c) the number of heads.}
        \label{fig:parameter_sensitivity}
\end{figure*}

We also investigate the sensitivity of parameters and report the results (AUC-ROC) of our proposed model with various parameters in Figure \ref{fig:parameter_sensitivity}. We consider the following three parameters:
\begin{itemize}
    \item \textit{Convolution hidden size dimension} ($H^m$): We first test the effect of the dimension of hidden layer in the subgraph convolution network. The results are shown in Figure \ref{fig:parameter_sensitivity}(a). We can observe that our model achieves the best performance when $H^m$ is set to 32. However, the performance is very sensitive with the change in the dimension, suggesting that the convolution hidden size $H^m$ plays a crucial role in comparison to other parameters.
    \item \textit{Attention hidden size dimension} ($\tilde{H}^m$): We explore our experimental results by varying the size of the attention vector ($\tilde{H}^m$) during projection. The result is shown in Figure \ref{fig:parameter_sensitivity}(b). We can find that the performance of \modelname\ generally slightly varies upto the hidden dimension size 128. For higher dimensions, the model struggles to converge and overfits as the 32-dimensional embedding is projected into a 256-dimensional embedding. 
    \item \textit{Number of heads} ($d$): In order to check the impact of our proposed aggregation framework, we measure the performance of \modelname\ with the number of heads $d$. The result is shown in Figure \ref{fig:parameter_sensitivity}(c). Based on the results, we observe that multi-head HSA model ($d>1$) performs better than single head ($d=1$). But as the complexity increases ($d=6$), the model starts to overfit, and the performance is significantly impacted. Also, we find that the training process becomes more stable with the growth in the number of model heads.
\end{itemize}

\subsection{Qualitative Analysis}
We qualitatively examine all 25 false negatives produced by our \modelname\ model in the testing phase of our dataset. In a Twitter network of collusive users, the priority is to minimize social damage; we, thereby, review the collusive users which are falsely detected as non-collusive by our model. We find that 18 of these users (\textasciitilde72\%) have less than 25 followers, thereby forming a sparse connectivity in our proposed heterogeneous graph. This also makes them unable to learn the connectivity information between the nodes and thereby limiting the performance of \modelname\ to identify these users. 
For the remaining 7 users, we could not find enough conclusive evidence from our collected data to interpret their incorrect detection by our proposed model. We would also like to mention that after manual investigation, 4 out of these 7 users do not exists currently on Twitter.

\begin{figure}[ht]
\centering
\begin{minipage}[b]{0.54\linewidth}
\centering
\includegraphics[width=0.9\textwidth]{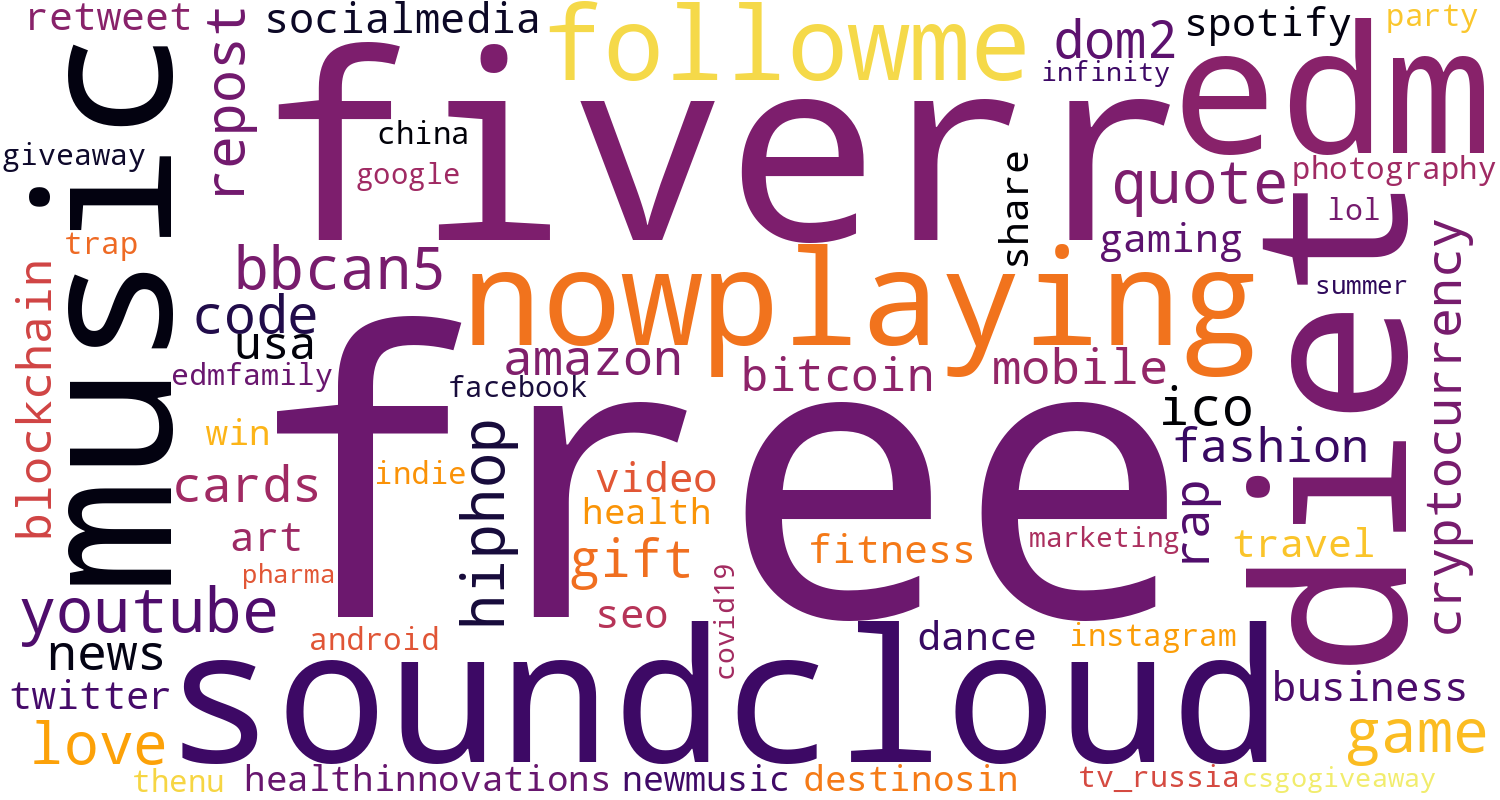}
\captionof{figure}{Wordcloud of the hashtags used by collusive users (true negatives)}
\label{fig:post_analysis}
\end{minipage}\hfill
\begin{minipage}[b]{0.42\linewidth}
\centering
\includegraphics[width=\textwidth]{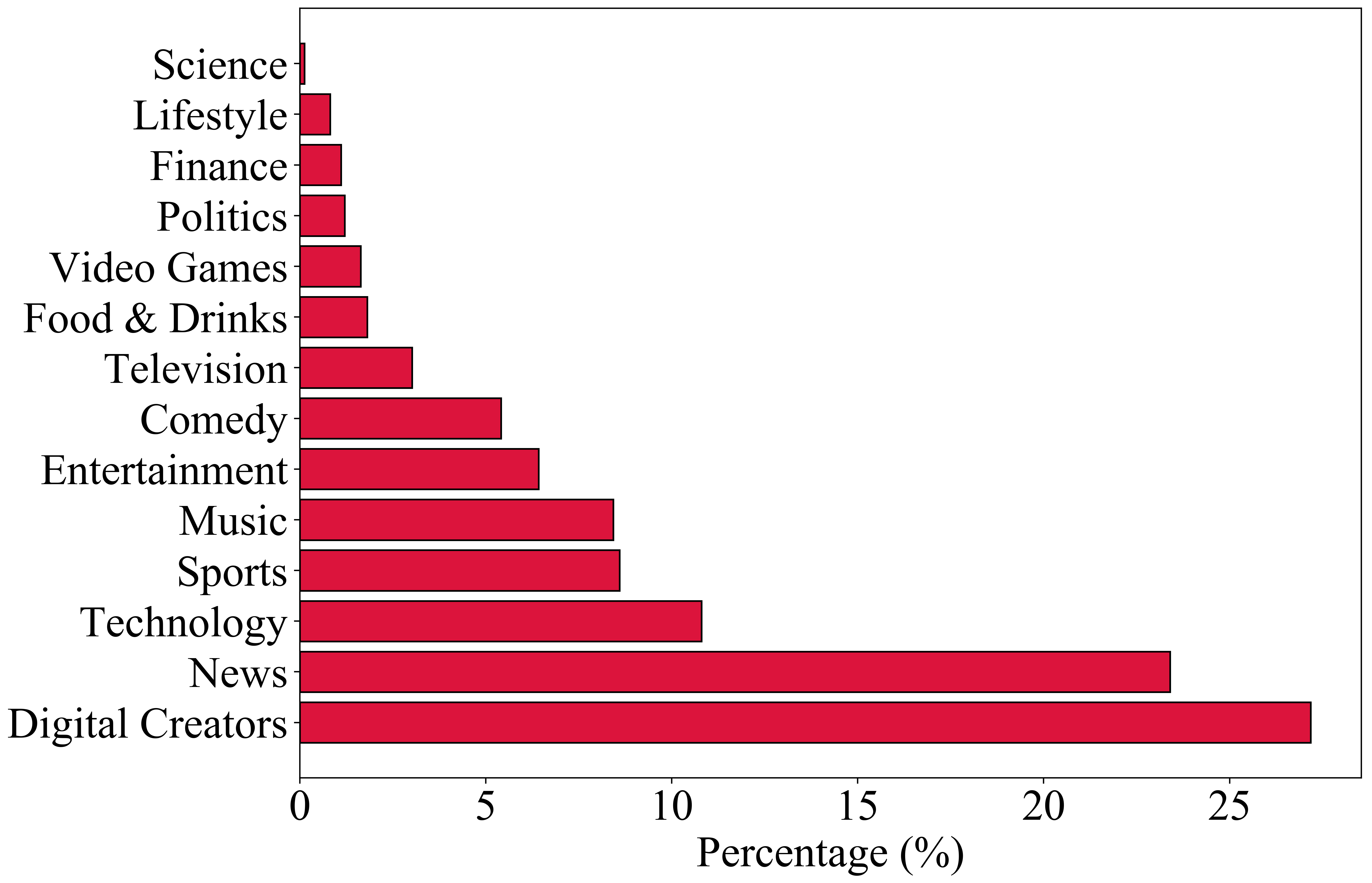}
\captionof{figure}{Category distribution of tweets posted by collusive users}
\label{fig:category_dist}
\end{minipage}
\end{figure}

In order to understand the behaviour of the correctly identified collusive users (true negatives\footnote{Note that collusive class is the negative class in our experiment.}), we highlight interesting quantitative covariates among these accounts. \\
\textbf{Current user status:} At the time of conducting this research, we found that only 2\% of detected collusive users which are collected from the blackmarket services are suspended/deleted by Twitter. This clearly shows how these users are successfully able to evade the existing fraudulent account detection techniques deployed by Twitter. \\
\textbf{Popular hashtags:} We analyze a total of 119,486 tweets posted by the collusive users detected using \modelname\ covering their entire tweeting activity in our dataset. Figure \ref{fig:post_analysis} shows the wordcloud of the hashtags used in tweets after removing two-letter words and common stopwords. Here, the font size corresponds to the frequency of the text. Unsurprisingly, we observe keywords related to the blackmarket services (`followme'), entertainment (`music', `soundcloud', `edm', `youtube', 'spotify'), freelance platforms(`fiverr'), digital asset (`cryptocurrency', `bitcoin', `blockchain'), etc. This shows how collusive entities could have manipulated the social growth of these entities. We believe these keywords also provide new directions for future collusive entity detection. We list out a few example tweets with these keywords in Table \ref{table:example_tweets}. \\
\textbf{Tweet categories:} It is often interesting to see which categories of tweets are posted by fraudulent users. We follow a zero-shot learning approach mentioned in \cite{yinroth2019zeroshot} to identify the categories present in the tweets posted by the detected collusive users. We frame it as a zero-shot topic-classification problem where the task is to map a sentence (tweet in our case) to a class (category in our case) that is not observed during training. We map the tweets into one of the 14 Interactive Advertising Bureau (IAB) categories\footnote{\url{https://developer.twitter.com/en/docs/twitter-ads-api/campaign-management/api-reference/iab-categories}} which is used by the supply partners within the Twitter Audience Platform (TAP). We show the category distribution of tweets posted by the detected collusive users in Figure \ref{fig:category_dist}. It can be clearly seen that almost 50\% of the tweets are from two categories: \textit{Digital creators} and \textit{News}. A possible reason for this is the recent popularity of these categories on Twitter, allowing tweets from these categories to connect directly with their audience, discover emerging content and trends, and build communities using two monetization programs\footnote{\url{https://media.twitter.com/en/articles/products/2018/media-studio/monetization.html}}: \textit{Amplify Pre-roll} and \textit{Amplify Sponsorships}.\\
\begin{table*}[!t]
\centering
\caption{Example of a  few  tweets posted by collusive users.}
    \label{table:example_tweets}
\begin{tabular}{ p{13cm}|p{2cm}}
    \hline
    \textbf{Tweet} & \textbf{Category} \\ \hline \hline
    Boost Your Google Ranking Fast with 250 High Quality Backlinks \#fiverr \#backlinks \#traffic \#website & Freelancing \\ \hline
    OMG, I have 1,000 followers! Thank you. Think I will make it to 10,000? \#followme & Blackmarkets \\ \hline
    This is the first spinner that brings free BTC. Spin \& Earn.  \#btcspinnerio \#btcspinner \#bitcoin \#btc & Digital asset \\ \hline
    This twitter site is all about \#sports \#gambling, sports \#handicapping. He give my free picks on \#Soccer, \#NFL-Foot & Sports\\ \hline
    Get up to 50GB of free space for all your photos, videos, docs, and music & Technology \\ \hline
    Check it out! socialspider will do viral youtube promotion usa or worldwide for \$5 on \#Fiverr & Freelancing \\ \hline
    \#CyberMiles is a new blockchain protocol being developed to revolutionize how digital commerce and online marketplace & Digital asset \\\hline
    Yo Bro my Music is poping right now. Everysong i drop on soundcloud touches 1000 in a Month. Can you drop my songs on your Youtube. & Entertainment \\ \hline
    Grab them while they last, 100 free DAT Tokens here \#datumnetwork \#ico \#cryptocurrency & Digital asset\\ \hline
    Get \$1million \$CPPG tokens. \#Airdrop \#bounty \#Crypto \#cryptocurrency & Digital asset\\ \hline
   \end{tabular}
\end{table*}

\section{Conclusion}
The prevalence of collusion over social media has become a serious  problem. In this paper, we addressed the problem of detecting collusive users involved in blackmarket services to produce artificial followers for Twitter users. We proposed \modelname, a framework that considers the follower/followee relationships and linguistic properties of tweets for detecting collusive users in Twitter. Compared with previous methods, our approach focuses on multiple  
relationships between the users and can leverage information more efficiently. Using extensive experiments, we demonstrated the effectiveness and superiority of \modelname\ with  an attention mechanism on our curated dataset over the state-of-the-art methods. 
To the best of our knowledge, this work is the first attempt using heterogeneous network to detect collusive followers on Twitter. 
The dataset we collected is also the first dataset of its kind. 
For future work, in addition to the metadata and topical properties, we plan to capture the temporal properties of these users that can help us  improve the performance of the predictive modeling task. We also plan to investigate the effectiveness of the proposed approach with closely related problems such as bot detection, identifying content polluters etc.
\section*{Acknowledgement}
The authors would like to acknowledge the support of ECR/2017/001691 (SERB) and ihub-Anubhuti-iiitd Foundation set up under the NM-ICPS scheme of the Department of Science and Technology, India.

\bibliographystyle{ACM-Reference-Format}
\bibliography{biblio}
≥
\end{document}